\documentclass[11pt,a4paper]{article}
\usepackage{jcappub}
\usepackage{subfigure}
\usepackage{url}

\renewcommand{\subfigcapskip}{5pt} 

\title{Gravitational waves from domain walls in the next-to-minimal supersymmetric standard model}

\author[a]{Kenji Kadota,}
\author[b,c]{Masahiro Kawasaki,}
\author[d]{and Ken'ichi Saikawa}

\affiliation[a]{Center for Theoretical Physics of the Universe,\\
Institute for Basic Science, Daejeon 305-811, Korea}
\affiliation[b]{Institute for Cosmic Ray Research, The University of Tokyo,\\
5-1-5 Kashiwa-no-ha, Kashiwa City, Chiba 277-8582, Japan} 
\affiliation[c]{Kavli Institute for the Physics and Mathematics of the Universe (WPI), Todai Institutes for Advanced Study, The University of Tokyo,\\
5-1-5 Kashiwa-no-ha, Kashiwa City, Chiba 277-8582, Japan}
\affiliation[d]{Department of Physics, Tokyo Institute of Technology,\\
2-12-1 Ookayama, Meguro-ku, Tokyo 152-8551, Japan}

\emailAdd{kadota@ibs.re.kr}
\emailAdd{kawasaki@icrr.u-tokyo.ac.jp}
\emailAdd{saikawa@th.phys.titech.ac.jp}
 
\abstract{The next-to-minimal supersymmetric standard model predicts the formation of domain walls due to the spontaneous breaking of the discrete $Z_3$-symmetry
at the electroweak phase transition, and they collapse before the epoch of big bang nucleosynthesis
if there exists a small bias term in the potential which explicitly breaks the discrete symmetry.
Signatures of gravitational waves produced from these unstable domain walls are estimated and their parameter dependence is investigated.
It is shown that the amplitude of gravitational waves becomes generically large in the decoupling limit, and that
their frequency is low enough to be probed in future pulsar timing observations.}


\begin{document}
\subheader{CTPU-15-03\\ IPMU15-0031}
\maketitle

%%%%%%%%%%%%%%%%%%%%%%%%%%%%%%%%%%%%%%%%%%%%%%%%%%%%%%%%%%%%%%%%
\section{\label{sec1} Introduction}
%%%%%%%%%%%%%%%%%%%%%%%%%%%%%%%%%%%%%%%%%%%%%%%%%%%%%%%%%%%%%%%%
Supersymmetry (SUSY) is a well-motivated extension of the standard model (SM) of particle physics
involving a solution to the hierarchy problem [see, e.g.,~\cite{Nilles:1983ge,Martin:1997ns} for reviews].
The simplest supersymmetric extension of the SM is called the minimal supersymmetric standard model (MSSM),
which consists of only SM particles along with a pair of Higgs doublets and their superpartners.
Such a model is also motivated by the unification of the gauge forces and the existence of dark matter candidates.
These interesting features will be investigated in various upcoming observations including high-energy physics,
astrophysics, and cosmology observable.

Aside from the rich phenomenology encountered in the MSSM, it suffers from the so-called $\mu$-problem~\cite{Kim:1983dt},
which is originated from the fact that the superpotential of the MSSM contains a term of the form $\mu H_uH_d$ (called the ``$\mu$-term") at the tree level.
Here $H_u$ and $H_d$ are two Higgs doublet superfields, and $\mu$ is some dimensionful parameter.
Since the MSSM might be a low energy effective theory derived from a more fundamental theory at higher energies,
we naively expect that the magnitude of the $\mu$ parameter is of the order of some cutoff scale such as the Planck scale.
On the other hand, $\mu$ should be of the order of the soft SUSY breaking scale in order to induce the electroweak symmetry breaking appropriately.
These considerations seem to revive the hierarchy problem in the MSSM.

The next-to-minimal supersymmetric standard model (NMSSM) [for reviews, see~\cite{Maniatis:2009re,Ellwanger:2009dp}]
is one of possible solutions to the $\mu$-problem, which introduces an additional singlet superfield
$S$ and replaces the $\mu$-term with the coupling of the form $\lambda SH_uH_d$.
Owing to this coupling, an appropriate magnitude of the $\mu$ parameter is dynamically generated when the scalar component of the singlet superfield
acquires a vacuum expectation value (VEV) at the electroweak phase transition.
However, the introduction of the singlet superfield allows the existence of other dimensionful parameters at the tree level.
The simplest way to prohibit such additional dimensionful parameters is to impose a discrete $Z_3$-symmetry,
and such a model is called the $Z_3$-invariant NMSSM, or often abridged as the NMSSM.

The discrete $Z_3$ symmetry of the NMSSM is spontaneously broken when the scalar components of the Higgs fields as well as the singlet scalar
acquire VEVs at the electroweak phase transition, which leads to the formation of domain walls in the universe.
The existence of such domain walls turns out to be problematic in the standard cosmology~\cite{Zeldovich:1974uw},
and hence it was argued that the NMSSM is unfavorable from the cosmological point of view~\cite{Abel:1995wk}.
Later, it was pointed out that there are some possibilities to arrange the model to have a small explicit symmetry breaking term,
which causes a late time annihilation of the walls~\cite{Panagiotakopoulos:1998yw,Hamaguchi:2011nm}.
Taking account of these possibilities, we deduce that unstable long-lived domain walls can exist in the early epoch of the universe.

In this paper, we point out that a stochastic background of gravitational waves with observably large amplitude
can be produced from domain walls in the NMSSM.
The amplitude of gravitational waves is determined once we specify the surface mass density of domain walls and their decay time~\cite{Hiramatsu:2010yz,Hiramatsu:2013qaa}.
Here we estimate the surface mass density by solving nonlinear field equations for the Higgs sector in the NMSSM to find a planar domain wall solution.
On the other hand, the peak frequency of gravitational waves solely depends on the decay time scale.
We see that the typical frequency of the gravitational waves corresponds to the band which can be probed via pulsar timing observations.
Another interesting finding is that the amplitude of gravitational waves becomes generically large in the so-called decoupling limit,
where the couplings between the singlet field and Higgs fields become extremely small.
Note that, due to the very weak coupling between the singlet-like states and the MSSM sector in the decoupling regime,
the signatures such as those from the collider tend to reduce to those in the MSSM, and we emphasize a complementary role of
the gravitational waves as a possible probe to distinguish the NMSSM from the MSSM.

Signatures of gravitational waves in the NMSSM were also investigated in Ref.~\cite{Apreda:2001tj} in the context of the first order electroweak phase transition.
Differently from that approach, in this paper we focus on the production of gravitational waves from domain walls, which does not depend on the order of the phase transition.
The crucial assumption is the occurrence of the spontaneous breaking of a discrete symmetry followed by the late time collapse of domain walls,
which is a definite consequence of the NMSSM with the (approximate) $Z_3$-symmetry.

The organization of this paper is as follows.
In Sec.~\ref{sec2}, we describe the scalar potential in the $Z_3$-invariant NMSSM, especially focusing on its Higgs sector.
In Sec.~\ref{sec3}, we introduce a numerical method to analyze the structure of domain walls, and discuss their cosmological evolution.
Then, in Sec.~\ref{sec4}, we estimate the amplitude and frequency of gravitational waves produced from domain walls and discuss prospects for future observations.
Finally, Sec.~\ref{sec5} is devoted to the conclusions.

%%%%%%%%%%%%%%%%%%%%%%%%%%%%%%%%%%%%%%%%%%%%%%%%%%%%%%%%%%%%%%%%
\section{\label{sec2} Higgs potential in the NMSSM}
%%%%%%%%%%%%%%%%%%%%%%%%%%%%%%%%%%%%%%%%%%%%%%%%%%%%%%%%%%%%%%%%
Let us consider the supersymmetric standard model with an additional gauge singlet superfield $S$.
As was described in Sec.~\ref{sec1}, we impose a discrete $Z_3$ symmetry, under which every chiral supermultiplet $\Phi$ transform as
\begin{equation}
\Phi \to e^{2\pi i/3}\Phi, \label{Z3_transf}
\end{equation}
in order to forbid the presence of dimensionful parameters.
The renormalizable superpotential of the NMSSM allowed by the $Z_3$ symmetry and the gauge invariance can be written as
\begin{equation}
W_{\rm NMSSM} = \lambda S H_uH_d + \frac{\kappa}{3}S^3 + W_{\rm Yukawa}, \label{W_NMSSM}
\end{equation}
where $\lambda$ and $\kappa$ are dimensionless couplings, and $W_{\rm Yukawa}$ corresponds to
the usual Yukawa couplings of the quark and lepton superfields of the MSSM.
In the above equations, contraction over gauge indices is understood.
For instance, we have
\begin{equation}
H_u = \left(
\begin{array}{c}
H_u^+\\
H_u^0
\end{array}
\right), \qquad
H_d = \left(
\begin{array}{c}
H_d^0\\
H_d^-
\end{array}
\right)
\end{equation}
for $SU(2)_L$ doublet fields, which leads
\begin{equation}
H_uH_d = H_u^+ H_d^- - H_u^0 H_d^0.
\end{equation}
Here and hereafter we use the same symbol for the superfield and its scalar component.
From F-term contributions of Eq.~\eqref{W_NMSSM} together with D-term contributions and soft SUSY breaking contributions, we obtain
Higgs potential at the tree level:
\begin{align}
V &= \left|\lambda(H_u^+H_d^- - H_u^0H_d^0)+\kappa S^2\right|^2 + (m_{H_u}^2 + |\lambda|^2|S|^2)(|H_u^0|^2 + |H_u^+|^2) \nonumber\\
&\quad +(m_{H_d}^2 + |\lambda|^2|S|^2)(|H_d^0|^2 + |H_d^-|^2) +\frac{g_1^2+g_2^2}{8}\left(|H_u^0|^2+|H_u^+|^2-|H_d^0|^2-|H_d^-|^2\right)^2 \nonumber\\
&\quad +\frac{g_2^2}{2}\left|H_u^+H_d^{0*}+H_u^0H_d^{-*}\right|^2 + m_S^2|S|^2 + \left[\lambda A_{\lambda}(H_u^+H_d^- - H_u^0H_d^0)S + \frac{1}{3}\kappa A_{\kappa}S^3 + \mathrm{h.c.} \right],
\end{align}
where $g_1$ and $g_2$ are the $U(1)_Y$ and $SU(2)_L$ gauge couplings, respectively, and $m_{H_u}^2$, $m_{H_d}^2$, $m_S^2$, $A_{\lambda}$,
and $A_{\kappa}$ are dimensionful soft parameters.

We can rotate away the component $H_u^+$ by using $SU(2)_L$ transformation, and hereafter we take $H_u^+=0$.
Furthermore, it is necessary to take $H_d^-=0$ in order to guarantee that the squared masses of charged components become positive.
Therefore, in the following we only consider the potential for the neutral components, and simply denote them as $H_u = H_u^0$ and $H_d = H_d^0$. The scalar potential becomes
\begin{align}
V &= \left|\kappa S^2 -\lambda H_uH_d\right|^2 + m_{H_u}^2|H_u|^2 + m_{H_d}^2|H_d|^2 + m_S^2|S|^2 + |\lambda|^2|S|^2(|H_u|^2 + |H_d|^2) \nonumber\\
&\quad+\frac{g^2}{4}\left(|H_u|^2-|H_d|^2\right)^2 + \left[\frac{1}{3}\kappa A_{\kappa}S^3 -\lambda A_{\lambda}H_uH_dS + \mathrm{h.c.} \right], \label{Higgs_potential}
\end{align}
where
\begin{equation}
g^2 \equiv \frac{g_1^2+g_2^2}{2} = \frac{2m_Z^2}{v^2} \simeq 0.275 \label{eq6}
\end{equation}
for the $Z$ boson mass $m_Z\simeq91.2\mathrm{GeV}$ and the Higgs VEV $v\simeq246\mathrm{GeV}$.
Here we assume that there is no explicit CP violation, and take $\lambda$, $\kappa$, $A_{\lambda}$, and $A_{\kappa}$ to be real.

It is possible to take the VEV of $\langle H_u\rangle$ to be real and positive
by the use of $U(1)_Y$ transformation.
In general, $\langle H_d\rangle$ and $\langle S \rangle$ are complex, but such an extremum becomes always a local maximum
if there is no explicit CP violation~\cite{Romao:1986jy}.
Therefore, we can also take $\langle H_d\rangle$ and $\langle S\rangle$ to be real up to the $Z_3$ transformations.\footnote{If we parameterize general three complex VEVs as
\begin{equation}
\langle H_u\rangle = \frac{1}{\sqrt{2}}v_ue^{i\phi_u}, \quad \langle H_d\rangle = \frac{1}{\sqrt{2}}v_de^{i\phi_d}, \quad \langle S\rangle = \frac{1}{\sqrt{2}}v_se^{i\phi_s},
\end{equation}
with $v_u$, $v_d$ and $v_s$ being real, relevant CP-violating phases are following two combinations:
\begin{equation}
\theta = \phi_u + \phi_d + \phi_s \quad \mathrm{and} \quad \delta = 3\phi_s. \label{CPV_phase}
\end{equation}
The no-go theorem of Ref.~\cite{Romao:1986jy} ensures that these two phases take trivial values at the minimum of the potential.
If we fix the $U(1)_Y$ gauge such that $v_u>0$ and $\phi_u=0$, this fact implies that there are two more vacua with $(\phi_d,\phi_s)=(2\pi/3,4\pi/3)$
and $(4\pi/3,2\pi/3)$ in addition to that given by Eq.~\eqref{vacuum_real}.
Note that it is also possible to choose the $U(1)_Y$ gauge such that these three degenerate vacua are represented by
$\phi_i=0$, $2\pi/3$, and $4\pi/3$ ($i=u,d,s$), respectively.}
Then, we parameterize three VEVs by using three real quantities:
\begin{equation}
\langle H_u\rangle = \frac{v_u}{\sqrt{2}},\quad \langle H_d\rangle = \frac{v_d}{\sqrt{2}}, \quad \langle S\rangle = \frac{v_s}{\sqrt{2}}, \label{vacuum_real}
\end{equation}
with $v_u>0$. Furthermore, we see that the potential possesses the following symmetries: $\lambda$, $\kappa$, $v_s$ $\to$ $-\lambda$, $-\kappa$, $-v_s$
and $\lambda$, $v_d$ $\to$ $-\lambda$, $-v_d$~\cite{Cerdeno:2004xw}.
Using these symmetries, we fix the signs of $\lambda$ and $v_d$ to be positive.
Other parameters $\kappa$, $v_s$, $A_{\lambda}$, and $A_{\kappa}$ can take both signs.

The stationary conditions of the potential $\partial V/\partial H_u = \partial V/ \partial H_d = \partial V/\partial S = 0$ lead to
\begin{align}
\lambda^2(v_s^2+v_d^2)v_u -\lambda\kappa v_dv_s^2 - \frac{g^2}{2}v_u(v_d^2-v_u^2) + 2m_{H_u}^2v_u - \sqrt{2}\lambda A_{\lambda}v_dv_s &= 0, \label{stationary1}\\
\lambda^2(v_s^2+v_u^2)v_d -\lambda\kappa v_uv_s^2 + \frac{g^2}{2}v_d(v_d^2-v_u^2) + 2m_{H_d}^2v_d - \sqrt{2}\lambda A_{\lambda}v_uv_s &= 0, \label{stationary2}\\
\lambda^2(v_u^2+v_d^2)v_s +2\kappa^2v_s^3 - 2\lambda\kappa v_uv_dv_s + 2m_S^2v_s - \sqrt{2}\lambda A_{\lambda}v_uv_d + \sqrt{2}\kappa A_{\kappa}v_s^2 &= 0. \label{stationary3}
\end{align}
Once we specify the values of $v_u$, $v_d$, and $v_s$, the soft masses $m_{H_u}^2$, $m_{H_d}^2$, and $m_S^2$ are determined from Eqs.~\eqref{stationary1}-\eqref{stationary3}:
\begin{align}
m_{H_u}^2 &= -\mu^2 - \frac{\lambda^2}{2}v^2\cos^2\beta + \frac{\kappa}{\lambda}\mu^2\cot\beta + \frac{g^2}{4}v^2\cos 2\beta + \mu A_{\lambda}\cot\beta,\\
m_{H_d}^2 &= -\mu^2 - \frac{\lambda^2}{2}v^2\sin^2\beta + \frac{\kappa}{\lambda}\mu^2\tan\beta - \frac{g^2}{4}v^2\cos 2\beta + \mu A_{\lambda}\tan\beta,\\
m_S^2 &= -\frac{\lambda^2}{2}v^2 - 2\frac{\kappa^2}{\lambda^2}\mu^2 + \frac{1}{2}\lambda\kappa v^2\sin 2\beta + \frac{\lambda^2}{4}\frac{A_{\lambda}v^2}{\mu}\sin 2\beta - \frac{\kappa}{\lambda}\mu A_{\kappa},
\end{align}
where
\begin{equation}
\mu \equiv \frac{\lambda v_s}{\sqrt{2}}, \quad \tan\beta \equiv \frac{v_u}{v_d}, \quad \mathrm{and} \quad v = \sqrt{v_u^2+v_d^2}. \label{mu_def}
\end{equation}
Hereafter, we use the following six quantities as free parameters of the model:
\begin{equation}
\lambda, \quad \kappa, \quad A_{\lambda}, \quad A_{\kappa}, \quad \tan\beta, \quad \mathrm{and} \quad \mu. \label{NMSSM_parameters}
\end{equation}

The definition of the $\mu$-parameter in Eq.~\eqref{mu_def} implies that the value of $v_s$ becomes much larger than the electroweak scale
for $\lambda\ll 1$ even though $\mu$ takes a value of $\mathcal{O}(100)$GeV.
Such a case is called the decoupling limit, since the coupling between particles associated with the singlet superfield and the MSSM sector
becomes negligible for $\lambda\to 0$.
In this limit, dominant terms in the potential~\eqref{Higgs_potential}
are given by 
\begin{equation}
V \simeq \kappa^2|S|^4 + m_S^2 |S|^2 + \left[\frac{1}{3}\kappa A_{\kappa}S^3 + \mathrm{h.c.} \right], \label{V_approx}
\end{equation}
from which we estimate the value of $v_s$ at the minimum as
\begin{equation}
v_s \simeq 
-\frac{\sqrt{2}A_{\kappa}}{4\kappa}\left(1+\sqrt{1-\frac{8m_S^2}{A_{\kappa}^2}}\right). \label{vs_approx}
\end{equation}
Note that the condition $A_{\kappa}^2\gtrsim 9m_S^2$ should be satisfied in order that
the extremum given by Eq.~\eqref{vs_approx} becomes deeper than the other possible minimum with $v_s=0$~\cite{Ellwanger:1996gw}.
Assuming that the magnitudes of $|A_{\kappa}|$ and $|m_S|$ are comparable to the soft SUSY breaking mass scale,
from Eq.~\eqref{vs_approx} we see that a large value of $v_s$ is compatible with $\kappa \ll1$.
Furthermore, it can be shown that $\kappa$ and $\lambda$ should be comparable such that the potential
does not have unrealistic minima~\cite{Ellwanger:1996gw,Kanehata:2011ei}.
Therefore, the decoupling limit is given by
\begin{equation}
\lambda\sim\kappa\to 0 \qquad \mathrm{and} \qquad v_s \sim -A_{\kappa}/\kappa \gg \mu, \label{decoupling_limit}
\end{equation}
with other dimensionful parameters being fixed.

We emphasize that the relation [Eq.~\eqref{decoupling_limit}] does not hold if $v_s$ exceeds some critical value,
since there might exist non-renormalizable operators which dominate over $\kappa^2|S|^4$ term in Eq.~\eqref{V_approx}
when we make the value of $\kappa$ sufficiently small. In this regime, the value of $v_s$ is not given by Eq.~\eqref{vs_approx}, 
but determined by minimizing the scalar potential including possible non-renormalizable terms.
Therefore, we expect that the relation [Eq.~\eqref{decoupling_limit}] remains valid only for some intermediate range of couplings,
and that the value of $v_s$ saturates to some finite value for a sufficiently small value of $\kappa$.

%%%%%%%%%%%%%%%%%%%%%%%%%%%%%%%%%%%%%%%%%%%%%%%%%%%%%%%%%%%%%%%%
\section{\label{sec3} Domain walls in the NMSSM}
%%%%%%%%%%%%%%%%%%%%%%%%%%%%%%%%%%%%%%%%%%%%%%%%%%%%%%%%%%%%%%%%
The $Z_3$-symmetry of the potential [Eq.~\eqref{Higgs_potential}] is spontaneously broken
when the scalar fields $H_u$, $H_d$ and $S$ acquire VEVs at the electroweak phase transition.
At that time, sheet-like objects called domain walls are created as a consequence of the spontaneous breaking of the discrete symmetry.
In this section, we describe the properties of domain walls created in the NMSSM and some theoretical issues related to them.
First, we analyze the domain wall solution to estimate the surface mass density in Sec.~\ref{sec3-1}.
Then, we discuss some cosmological aspects of NMSSM domain walls in Sec.~\ref{sec3-2}.
%%%%%%%%%%%%%%%%%%%%%%%%%%%%%%%%%%%%%%%%%%%%%%%%%%%%%%%%%%%%%%%%
\subsection{\label{sec3-1} Structure of domain walls}
%%%%%%%%%%%%%%%%%%%%%%%%%%%%%%%%%%%%%%%%%%%%%%%%%%%%%%%%%%%%%%%%
Since the potential is invariant under the $Z_3$ transformation [Eq.~\eqref{Z3_transf}],
there exist three degenerate vacua parameterized by ($\langle S\rangle$, $\langle H_u\rangle$, $\langle H_d\rangle$) $=$ ($v_s/\sqrt{2}$, $v_u/\sqrt{2}$, $v_d/\sqrt{2}$), 
($v_s e^{i2\pi/3}/\sqrt{2}$, $v_u e^{i2\pi/3}/\sqrt{2}$, $v_d e^{i2\pi/3}/\sqrt{2}$), and ($v_s e^{i4\pi/3}/\sqrt{2}$, $v_u e^{i4\pi/3}/\sqrt{2}$, $v_d e^{i4\pi/3}/\sqrt{2}$).
Domain walls are located around boundaries of these three vacua.

In order to find the domain wall solution [i.e., find the spatial configuration of the scalar fields $(S({\bf x}),H_u({\bf x}),H_d({\bf x}))$ around the core of the domain wall],
we must solve nonlinear field equations for three complex scalar fields $S$, $H_u$ and $H_d$ with the potential given by Eq.~\eqref{Higgs_potential}.
It is straightforward to solve such a system numerically~\cite{Abel:1995uc,Abel:1995wk}:
A planar domain wall solution $(S(z),H_u(z),H_d(z))$ perpendicular to the $z$-axis can be found from the following equations
\begin{equation}
\frac{d^2S}{dz^2} = \frac{\partial V}{\partial S^*}, \quad \frac{d^2H_u}{dz^2} = \frac{\partial V}{\partial H_u^*}, \quad \mathrm{and}\quad \frac{d^2H_d}{dz^2} = \frac{\partial V}{\partial H_d^*},
\label{EOM_numerical} 
\end{equation}
with boundary conditions 
\begin{align}
(S(z),H_u(z),H_d(z)) &\xrightarrow{z\to -\infty} \left(\frac{v_s}{\sqrt{2}},\frac{v_u}{\sqrt{2}},\frac{v_d}{\sqrt{2}}\right), \nonumber\\
(S(z),H_u(z),H_d(z)) &\xrightarrow{z\to +\infty} \left(\frac{v_se^{i\frac{2\pi}{3}}}{\sqrt{2}},\frac{v_ue^{i\frac{2\pi}{3}}}{\sqrt{2}},\frac{v_de^{i\frac{2\pi}{3}}}{\sqrt{2}}\right).
\end{align}
Here we use the globally convergent Newton method~\cite{Press:2007zz} to solve this boundary value problem.
Namely, we first specify the ``initial guess" for $(S(z),H_u(z),H_d(z))$ within some finite interval $-L\le z\le L$,
and deform it such that Eq.~\eqref{EOM_numerical} is satisfied with an appropriate boundary conditions specified at $z=\pm L$.
Then, we iterate this procedure until Eq.~\eqref{EOM_numerical} is satisfied with an accuracy of $\mathcal{O}(10^{-5})$.
For an appropriate choice of the initial guess of the functions $(S(z),H_u(z),H_d(z))$,
this algorithm converges after a few iteration steps.

\begin{figure*}[htbp]
\def\subfigcapskip{8pt}
\centering
$\begin{array}{cc}
\subfigure[]{
\includegraphics[width=0.45\textwidth]{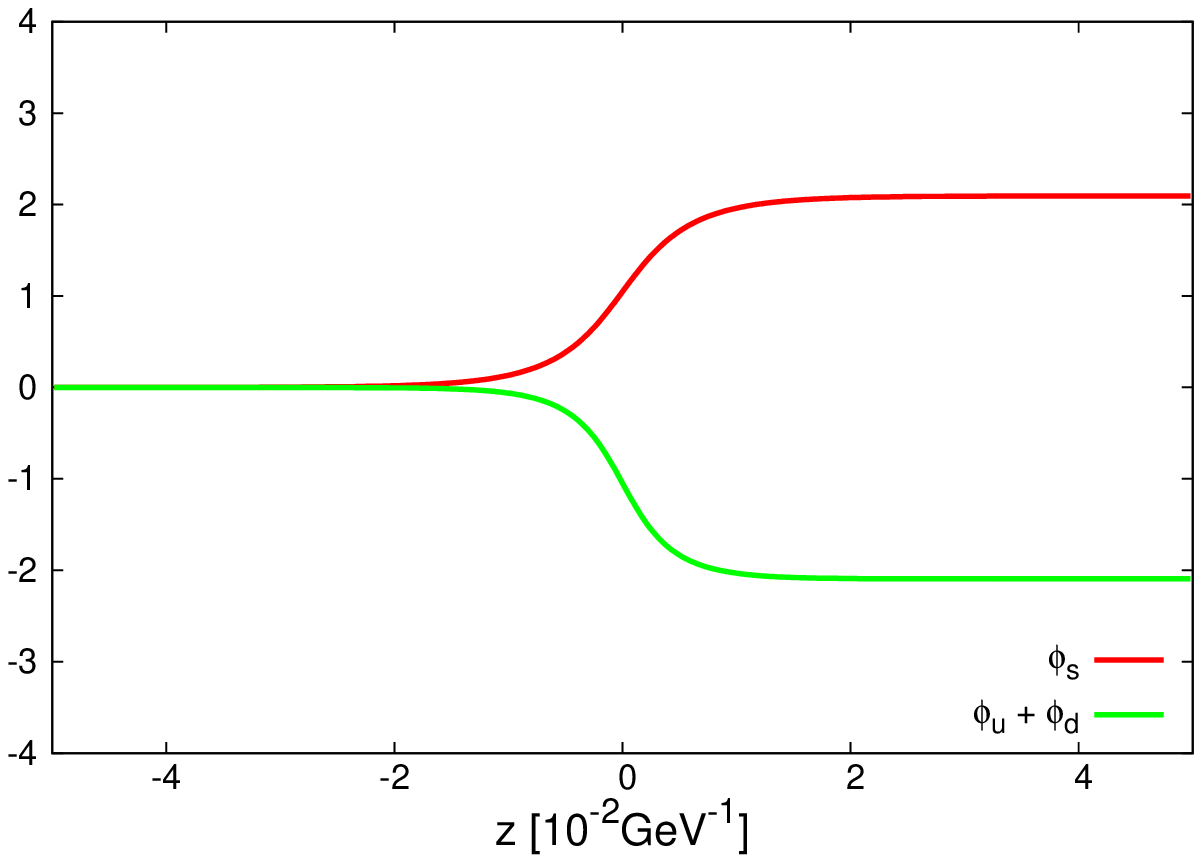}}
\hspace{20pt}
\subfigure[]{
\includegraphics[width=0.45\textwidth]{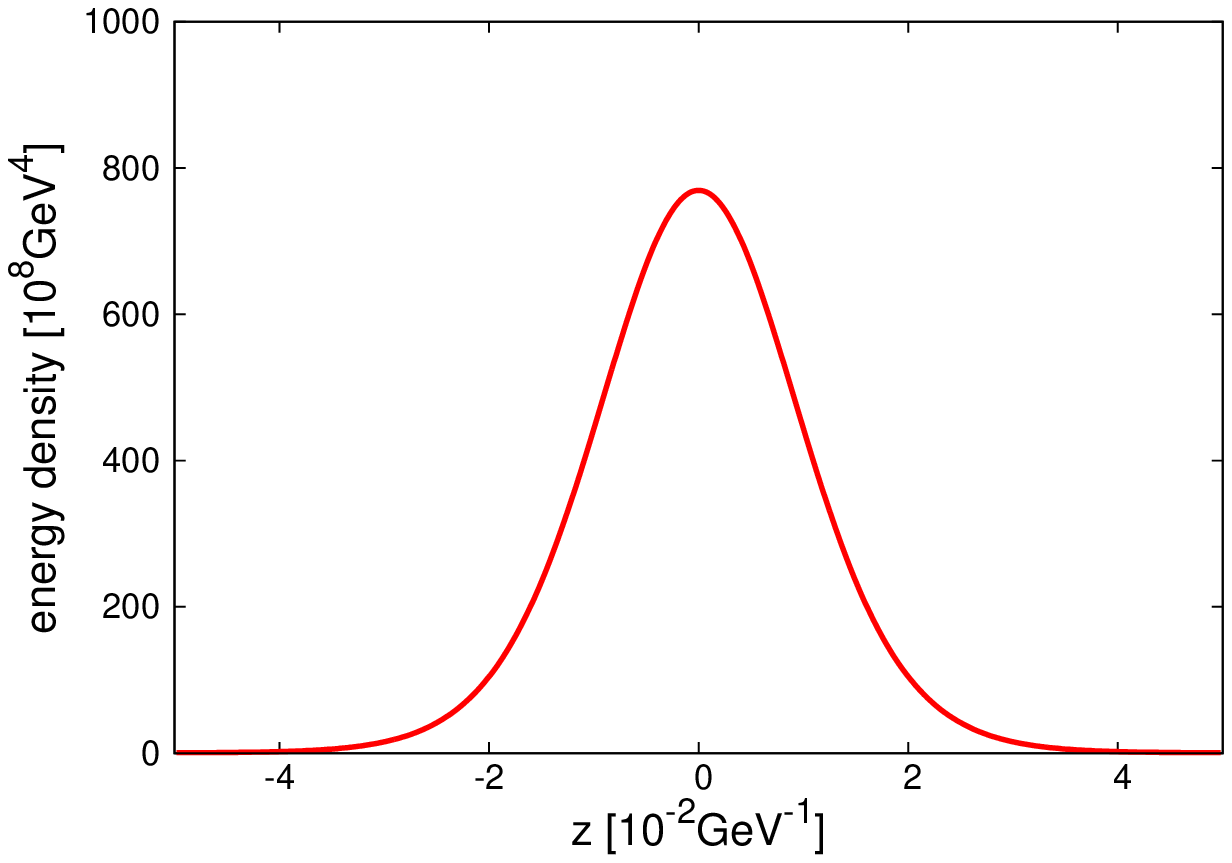}}
\end{array}$
\caption{Spatial distribution of (a) phases and (b) energy density of scalar fields $(S(z),H_u(z),H_d(z))$ of the planer domain wall solution
for the parameters $\lambda=0.05$, $\kappa=0.02$, $A_{\lambda}=150$GeV, $A_{\kappa}=-150$GeV, $\tan\beta=5$ and $\mu=200$GeV.
In the left panel, $\phi_s$, $\phi_u$ and $\phi_d$ represent the phases of $S$, $H_u$ and $H_d$, respectively.
Here we just show the sum of $\phi_u$ and $\phi_d$, since the relative phase between $H_u$ and $H_d$ suffer from
an arbitrariness due to the $U(1)_Y$ gauge transformations.
}
\label{fig1}
\end{figure*}

Figure~\ref{fig1} shows a typical configuration of the domain wall solution.
From Fig.~\ref{fig1} (a) we see that the phases of scalar fields continuously change around the core of the domain walls centered at $z=0$.
We also compute the spatial distribution of the energy density for this solution
\begin{align}
\rho_{\rm wall}(z) &= \left|\frac{d S}{dz}(z)\right|^2 + \left|\frac{d H_u}{dz}(z)\right|^2 + \left|\frac{d H_d}{dz}(z)\right|^2 \nonumber\\
&\quad+ V(S(z),H_u(z),H_d(z)) - V(v_s/\sqrt{2},v_u/\sqrt{2},v_d/\sqrt{2}),\label{rho_wall}
\end{align}
where we subtracted a constant term such that $\rho_{\rm wall} \to 0$ is satisfied for $z\to \pm\infty$.
As shown in Fig.~\ref{fig1} (b), the energy of the scalar fields is concentrated around the core of the domain wall.
Finally, we estimate the surface mass density $\sigma_{\rm wall}$ of the domain wall by integrating this
energy density along the z-axis:
\begin{equation}
\sigma_{\rm wall} = \int dz\rho_{\rm wall}(z).
\end{equation}
For a choice of parameters used in Fig.~\ref{fig1}, we obtain $\sigma_{\rm wall}\simeq 1.9\times 10^9\mathrm{GeV}^3$.

Although the precise value of $\sigma_{\rm wall}$ can only be calculated by using the numerical procedure,
we can roughly estimate the parameter dependence of $\sigma_{\rm wall}$ in the simple decoupling limit given by Eq.~\eqref{decoupling_limit}.
Regarding the fact that the phase of the $S$ field continuously changes across the core of the wall and that the potential can be approximated as Eq.~\eqref{V_approx}
in the decoupling limit, we expect that the typical length scale of the spatial variation of the field $S$ around the core of the wall is given by $\Delta z \sim |\kappa A_{\kappa}v_s|^{-1/2}$.
Furthermore, the estimation shown in Eq.~\eqref{vs_approx} implies that all three terms in the right-hand side of Eq.~\eqref{V_approx} are comparable,
and hence we roughly estimate the height of the potential around the core of the wall as $V\sim|\kappa A_{\kappa}v_s^3|$.
These observations lead to the following scaling behavior of the surface mass density of domain walls:
\begin{equation}
\sigma_{\rm wall} \sim \kappa v_s^3. \label{sigma_decoupling_limit}
\end{equation}
In fact, we confirmed that this behavior holds very accurately for $\lambda$, $\kappa < \mathcal{O}(10^{-4})$ in our numerical results.
Eq.~\eqref{sigma_decoupling_limit} indicates that the surface mass density of domain walls becomes enormously large in the decoupling limit.
We will discuss the observational consequence of this property in Sec.~\ref{sec4}.

%%%%%%%%%%%%%%%%%%%%%%%%%%%%%%%%%%%%%%%%%%%%%%%%%%%%%%%%%%%%%%%%
\subsection{\label{sec3-2} Domain wall problem and its solution}
%%%%%%%%%%%%%%%%%%%%%%%%%%%%%%%%%%%%%%%%%%%%%%%%%%%%%%%%%%%%%%%%
Let us turn our attention to the cosmological evolution of domain walls.
After the formation of domain walls, two kinds of forces act on them.
One is the tension force, which smooths out small scale structures to straighten the wall, and the other is the friction force caused by the interactions
between particles in the thermal plasma and Higgs fields constituting the wall.
The friction force is efficiently induced by particles whose masses are comparable to the electroweak scale, since their interactions with the Higgs fields are relatively strong.
However, the number density of these particles drops exponentially as the temperature of the plasma $T$ decreases,
and eventually the friction force becomes negligible. The friction effect for the NMSSM domain walls was quantitatively estimated in Ref.~\cite{Abel:1995wk},
and it was shown that this effect becomes irrelevant for $T<\mathcal{O}(0.1\textendash 1)$GeV.

Once the friction force becomes negligible, the dynamics of domain walls is mostly determined by the tension force,
which acts like $p_T \sim \sigma_{\rm wall}/R_{\rm wall}$ for a patch of the walls whose typical curvature radius is given by $R_{\rm wall}$.
In this case, it is known that the evolution of domain walls is described by the so-called scaling solution,
in which the typical length scales of the system are given by the Hubble radius.
In particular, the energy density of domain walls in the scaling regime is given by~\cite{Hiramatsu:2013qaa}
\begin{equation}
\rho_{\rm wall} = \mathcal{A}\frac{\sigma_{\rm wall}}{t}, \label{rho_wall_saling}
\end{equation}
where $\mathcal{A}$ is a dimensionless coefficient called the area parameter.
By comparing a naive estimation $\rho_{\rm wall}\sim \sigma_{\rm wall}/R_{\rm wall}$ with Eq.~\eqref{rho_wall_saling}, we estimate the typical curvature radius of the walls in
the scaling regime as $R_{\rm wall}\sim t/\mathcal{A}$, and hence the tension force is given by $p_T\sim \mathcal{A}\sigma_{\rm wall}/t$.

In the literature, the scaling behavior of domain walls [Eq.~\eqref{rho_wall_saling}] is confirmed both numerically and analytically~\cite{Press:1989yh,Hiramatsu:2013qaa}.
The result of the numerical simulation for the model of a real scalar field with $Z_2$ symmetry
indicates ${\cal A}\simeq 0.8\pm0.1$~\cite{Hiramatsu:2013qaa}, while other simulations for the model with $N$ degenerate vacua
show that the value of ${\cal A}$ increases proportionally with $N$~\cite{Hiramatsu:2012sc}.
Since we are interested in the model with three degenerate vacua (i.e. $Z_3$ invariant NMSSM), we simply estimate the value of ${\cal A}$ as ${\cal A}\simeq 0.8\times(3/2) = 1.2$.

If the domain walls evolve according to the scaling solution,
the decrement of the energy density $\rho_{\rm wall}\propto 1/t$ is slower than that of dusts $\propto a(t)^{-3}$ and radiations $\propto a(t)^{-4}$,
where $a(t)$ is the scale factor of the universe.
This fact implies that domain walls are likely to dominate the energy density of the universe at later times.
Such an existence of domain walls leads to many problems in the standard cosmological scenario~\cite{Zeldovich:1974uw}.

A simple solution to the domain wall problem is to introduce an additional term $\Delta V$
which explicitly breaks the discrete symmetry in the potential~\cite{Zeldovich:1974uw,Vilenkin:1981zs}.
Let us call this additional term a {\it bias}.
We note that the magnitude of the bias term should be much less than that of the potential around the core of domain walls ($\Delta V\ll V$)
such that the discrete $Z_3$-symmetry holds approximately.
If such a bias term exists in the potential, it breaks the degeneracy between different vacua, and the energy difference between
neighboring domains causes the volume pressure $p_V\sim \Delta V$ acting on the walls.
Domain walls start to collapse when this pressure $p_V$ exceeds the tension force $p_T\sim \mathcal{A}\sigma_{\rm wall}/t$,
and we can estimate their decay time $t_{\rm dec}$ from the condition $p_V\simeq p_T$:
\begin{equation}
t_{\rm dec} \simeq \frac{\mathcal{A}\sigma_{\rm wall}}{\Delta V}. \label{t_dec}
\end{equation}
If the decay of domain walls occurs before they overclose the universe, we have $\rho_{\rm wall}(t_{\rm dec})<\rho_{\rm crit}(t_{\rm dec})=3H^2(t_{\rm dec})/8\pi G$,
where $\rho_{\rm crit}(t_{\rm dec})$ is the critical density of the universe at the time $t_{\rm dec}$, $H(t_{\rm dec})$ is the Hubble parameter at that time,
and $G$ is the Newton's constant. From this requirement, we obtain
\begin{equation}
\sigma_{\rm wall} < 2.93\times 10^4\mathrm{TeV}^3\ \mathcal{A}^{-1}\left(\frac{0.1\mathrm{sec}}{t_{\rm dec}}\right). \label{condition_avoid_WD}
\end{equation}

Even if domain walls decay before they overclose the universe, the consideration of big bang nucleosynthesis (BBN) leads to another constraint if they are sufficiently long-lived.
For domain walls existing around the epoch of BBN, the ratio between their energy density and the entropy density $s$ is given by
\begin{equation}
\frac{\rho_{\rm wall}}{s} \simeq 2\times 10^{-7}\mathrm{GeV}\ \mathcal{A}\left(\frac{\sigma_{\rm wall}}{1\mathrm{TeV}^3}\right)\left(\frac{t}{1\mathrm{sec}}\right)^{1/2}. \label{rho_wall_to_s}
\end{equation}
It is expected that particles radiated from the walls are likely to destroy light elements created during the epoch of BBN.
For the abundance of radiated particles, which is comparable with Eq.~\eqref{rho_wall_to_s}, we require $t_{\rm dec}\lesssim0.01\mathrm{sec}$ in order to avoid
the constraints on hadronic energy injection in the standard BBN scenario~\cite{Kawasaki:2004yh}.
This condition gives a lower limit on the magnitude of the bias term:
\begin{equation}
\Delta V \gtrsim 6.6\times 10^{-2}\mathrm{MeV}^4\ \mathcal{A}\left(\frac{\sigma_{\rm wall}}{1\mathrm{TeV}^3}\right). \label{bias_lower_limit}
\end{equation}

Several remarks on the origin of the bias term [Eq.~\eqref{bias_lower_limit}] were made in the literature.
A naive expectation is that the discrete symmetry is slightly broken due to gravity, and that
a Planck-suppressed higher dimensional operator is responsible for the bias term~\cite{Rai:1992xw}.
However, it turned out that this possibility is unfavorable in the NMSSM~\cite{Abel:1995wk}, since
such a Planck-suppressed interaction radiatively induces a tadpole operator~\cite{Ellwanger:1983mg},
which revives the hierarchy problem because of the large VEV of the singlet field.
A solution to this problem was proposed in Ref.~\cite{Panagiotakopoulos:1998yw}:
It was pointed out that a small tadpole term of the form
\begin{equation}
\Delta V \sim \xi m_{\rm soft}^3S + \mathrm{h.c.} \label{bias_tadpole}
\end{equation}
can be generated by imposing various additional symmetries to constrain the form of nonrenormalizable interactions.
Here, $m_{\rm soft}$ is the soft SUSY breaking mass,
and the dimensionless coefficient $\xi$ corresponds to the loop suppression factor.
The term given by Eq.~\eqref{bias_tadpole} is small enough not to destabilize the singlet field, and it acts
as a bias term to remove domain walls.\footnote{We note that there are other various attempts
to avoid tadpole and domain wall problem~\cite{Abel:1996cr,Kolda:1998rm,Panagiotakopoulos:1999ah}.
These models also introduce additional symmetries to forbid a large tadpole term.
Since the form of the Lagrangian densities of these models becomes different from that of
the $Z_3$-invariant NMSSM [Eq.~\eqref{W_NMSSM}], which does not lead to the formation of domain walls,
we do not consider these possibilities in this paper.}

In addition to the solution described above, there is
another possibility, where a coupling between the singlet field and additional vector-like matters which are
charged under the SM $SU(3)_C$ gauge group or a {\it hidden} QCD gauge group is introduced~\cite{Hamaguchi:2011nm}.
In the presence of such a coupling, the $Z_3$-symmetry becomes anomalous for the QCD, and the bias term which destabilizes
the domain walls is generated due to the instanton effect~\cite{Preskill:1991kd}.
In this case, the magnitude of the bias term is estimated as
\begin{equation}
\Delta V \sim \Lambda^4, \label{bias_QCD}
\end{equation}
where $\Lambda$ corresponds to the QCD scale $\Lambda_{\rm QCD}\approx\mathcal{O}(100)\mathrm{MeV}$
if vector-like exotics are charged under $SU(3)_C$, or a scale at which the hidden gauge force becomes strong
if they are charged under the hidden gauge group.

Once the magnitude of the bias term is specified as Eq.~\eqref{bias_tadpole} or Eq.~\eqref{bias_QCD},
we can estimate the decay time of domain walls according to Eq.~\eqref{t_dec}.
Since the origin of the bias term is highly model dependent, 
we take $t_{\rm dec}$ as a free parameter in the rest of this paper.

%%%%%%%%%%%%%%%%%%%%%%%%%%%%%%%%%%%%%%%%%%%%%%%%%%%%%%%%%%%%%%%%
\section{\label{sec4} Gravitational waves from domain walls}
%%%%%%%%%%%%%%%%%%%%%%%%%%%%%%%%%%%%%%%%%%%%%%%%%%%%%%%%%%%%%%%%
From the discussion in the previous section, we see that domain walls exist after the electroweak phase transition, and that
they collapse before the epoch of BBN $t_{\rm dec}\lesssim 0.01\mathrm{sec}$ due to the effect of the bias term.
In this section, we estimate gravitational wave signatures produced
from these unstable domain walls~\cite{Gleiser:1998na,Hiramatsu:2010yz,Hiramatsu:2013qaa}.

The cosmological evolution of domain wall networks is accompanied by various violent processes of
collisions and annihilations of them. During these processes, a part of the energy stored in them is
released as gravitational waves, which leads to a stochastic gravitational wave background observed today.
The amplitude of gravitational waves at the cosmic time $t$ for a certain frequency $f$ can be characterized by the following quantity
\begin{equation}
\Omega_{\rm gw}(t) = \frac{1}{\rho_{\rm crit}(t)}\frac{d\rho_{\rm gw}(t)}{d\ln f},
\end{equation}
where $\rho_{\rm crit}(t)$ and $\rho_{\rm gw}(t)$ are the critical density of the universe at the time $t$ and the energy density of gravitational waves at that time, respectively.

The spectrum of gravitational waves from domain walls is estimated in Refs.~\cite{Hiramatsu:2010yz,Hiramatsu:2013qaa}
by the use of 3-dimensional lattice simulation of the scalar field.
According to the recent detailed study~\cite{Hiramatsu:2013qaa}, the spectrum 
of gravitational waves from domain walls has a peak at the frequency corresponding to the Hubble radius at the time $t_{\rm dec}$.
The peak amplitude of the gravitational waves at that time is given by~\cite{Hiramatsu:2013qaa}\footnote{In addition to the contribution from the dynamics of domain walls, there exists a secondary background due to the effort of the scalar fields to maintain the scaling defect~\cite{Figueroa:2012kw}. The spectrum of such a secondary background turns out to be scale invariant, extending over a wider frequency range. The amplitude of the scale-invariant background $\Omega_{\rm gw}^{(s)}$ can be estimated from the VEV of the scalar field, and in the NMSSM case we obtain $\Omega_{\rm gw}^{(s)}(t_{\rm dec}) \sim (v_s/M_{\rm Pl})^4$, where $M_P\simeq 1.22\times 10^{19}\mathrm{GeV}$ is the Planck mass. Since this amplitude is smaller than Eq.~\eqref{Omega_gw_t_dec} by a factor of $\sim (H(t_{\rm dec})/A_{\kappa})^2$, hereafter we do not consider this secondary contribution.}
\begin{align}
\Omega_{\rm gw}(t_{\rm dec})_{\rm peak} = \frac{8\pi\tilde{\epsilon}_{\rm gw}G^2{\cal A}^2\sigma_{\rm wall}^2}{3H^2(t_{\rm dec})}, \label{Omega_gw_t_dec}
\end{align}
where $\tilde{\epsilon}_{\rm gw}$ is an efficiency parameter of the gravitational radiation.
The result of the numerical simulation gives $\tilde{\epsilon}_{\rm gw}\simeq 0.7\pm0.4$~\cite{Hiramatsu:2013qaa}. Assuming that the gravitational radiation is terminated in
the radiation dominated era, we estimate the amplitude of the gravitational waves at the present time $t_0$ as
\begin{align}
\Omega_{\rm gw}h^2(t_0) &\simeq \Omega_Rh^2
\left(\frac{g_{S0}}{g_0}\right)^{4/3}
\left(\frac{g_0}{g_*}\right)^{1/3}\Omega_{\rm gw}(t_{\rm dec}) \nonumber\\
&\simeq 3.45\times 10^{-5}\times\left(\frac{10.75}{g_*}\right)^{1/3}\Omega_{\rm gw}(t_{\rm dec}), \label{Omega_gw_t0}
\end{align}
where $\Omega_Rh^2=4.15\times 10^{-5}$ is the density parameter of radiations at the present time, $h$ is the present value of the Hubble parameter in the unit of 
100km$\cdot$sec$^{-1}$Mpc$^{-1}$, 
$g_{S0}=3.91$ is the effective degrees of freedom for the entropy density at the present time,
and $g_0=3.36$ and $g_*$ are the number of radiation degrees of freedom at the present time and $t_{\rm dec}$, respectively.
Substituting Eqs.~\eqref{t_dec} and~\eqref{Omega_gw_t_dec} into Eq.~\eqref{Omega_gw_t0}, we obtain
\begin{align}
\Omega_{\rm gw}h^2(t_0)_{\rm peak} &\simeq 5.20\times 10^{-20}\times \tilde{\epsilon}_{\rm gw}{\cal A}^4\left(\frac{10.75}{g_*}\right)^{1/3}\left(\frac{\sigma_{\rm wall}}{1\mathrm{TeV}^3}\right)^4\left(\frac{1\mathrm{MeV}^4}{\Delta V}\right)^2. \label{Omega_gw_t0_peak}
\end{align}
The peak frequency is given by the Hubble parameter at the decay time:
\begin{align}
f(t_0)_{\rm peak} &= \frac{a(t_{\rm dec})}{a(t_0)}H(t_{\rm dec}) \nonumber\\
&\simeq 3.99\times 10^{-9} \mathrm{Hz}\ {\cal A}^{-1/2}\left(\frac{1\mathrm{TeV}^3}{\sigma_{\rm wall}}\right)^{1/2}\left(\frac{\Delta V}{1\mathrm{MeV}^4}\right)^{1/2}. \label{f_t0_peak}
\end{align}

We see that $\Omega_{\rm gw}h^2(t_0)_{\rm peak}$ and $f(t_0)_{\rm peak}$ are determined by two parameters: $\sigma_{\rm wall}$ and $\Delta V$
[or those appearing in the explicit from of the bias term, i.e., Eq.~\eqref{bias_tadpole} or Eq.~\eqref{bias_QCD}].
Furthermore, $\Delta V$ is related to the decay time of domain walls via Eq.~\eqref{t_dec},
and hence Eqs.~\eqref{Omega_gw_t0_peak} and~\eqref{f_t0_peak} are rephrased as
\begin{align}
\Omega_{\rm gw}h^2(t_0)_{\rm peak} &\simeq 1.20\times 10^{-17}\times \tilde{\epsilon}_{\rm gw}{\cal A}^4\left(\frac{10.75}{g_*}\right)^{1/3}\left(\frac{\sigma_{\rm wall}}{1\mathrm{TeV}^3}\right)^2\left(\frac{t_{\rm dec}}{0.01\mathrm{sec}}\right)^2, \label{Omega_gw_t0_peak_2} \\
f(t_0)_{\rm peak} &\simeq 1.02\times 10^{-9} \mathrm{Hz}\left(\frac{0.01\mathrm{sec}}{t_{\rm dec}}\right)^{1/2}. \label{f_t0_peak_2}
\end{align}
As was noted in Sec.~\ref{sec3-2}, here we use $t_{\rm dec}$ as a free parameter instead of $\Delta V$.

\begin{figure*}[htbp]
\begin{center}
\includegraphics[scale=0.8]{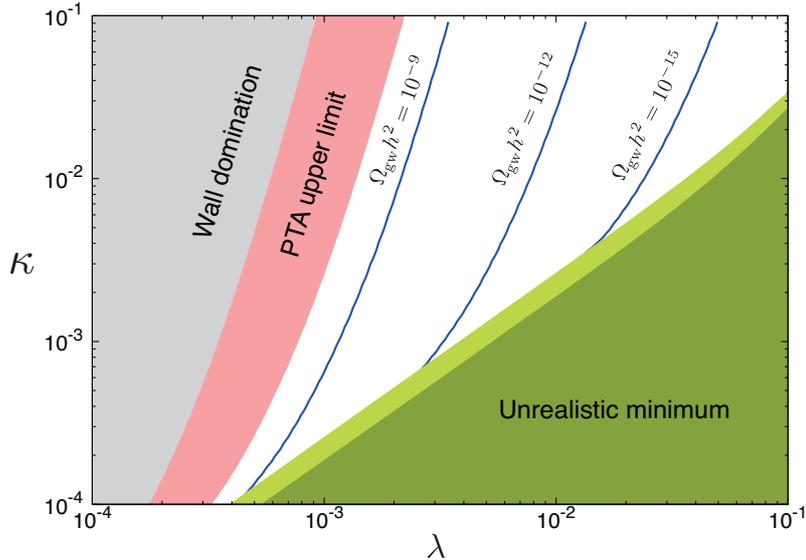}
\end{center}
\caption{The peak amplitude of gravitational waves estimated from Eq.~\eqref{Omega_gw_t0_peak_2} with the assumption of $t_{\rm dec}=0.01\mathrm{sec}$ in the $(\lambda,\kappa)$ plane.
Blue solid lines correspond to contours for $\Omega_{\rm gw}h^2(t_0)_{\rm peak}=10^{-9}$, $10^{-12}$, and $10^{-15}$.
Here we use $g_*=10.75$, $\mathcal{A}=1.2$ and $\tilde{\epsilon}_{\mathrm{gw}}=0.7$ for numerical coefficients appearing in Eq.~\eqref{Omega_gw_t0_peak_2}.
Other theoretical parameters are fixed as $A_{\lambda}=150$GeV, $A_{\kappa}=-150$GeV, $\tan\beta=5$ and $\mu=200$GeV.
The dark green region corresponds to the parameters at which the mass matrix of Higgs scalars has some negative eigenvalues,
and the light green region corresponds to the parameters at which
some unrealistic minimum becomes deeper than the desired minimum.
The pink region corresponds to the parameter space where the amplitude of gravitational waves exceeds the present limit $\Omega_{\rm gw}h^2<10^{-8}$ from PTA observations.
The gray region corresponds to the parameter space where domain walls dominate the energy density of the universe before $t_{\rm dec}$ [i.e. Eq.~\eqref{condition_avoid_WD}].}
\label{fig2}
\end{figure*}

In order to estimate the peak amplitude for a given set of the NMSSM parameters [Eq.~\eqref{NMSSM_parameters}],
we must evaluate the value of the surface mass density of domain walls $\sigma_{\rm wall}$ appearing in Eq.~\eqref{Omega_gw_t0_peak_2}.
Here, we use the numerical method introduced in Sec.~\ref{sec3-1} to estimate $\sigma_{\rm wall}$.
Figure~\ref{fig2} shows the estimated peak amplitude in two-dimensional parameter space of $(\lambda,\kappa)$.
In this figure, we also plotted some constraints on the tree-level Higgs potential:
One is the condition that the mass matrix of seven Higgs scalars (three CP-even, two CP-odd, and a pair of charged states)
do not have any negative eigenvalues, and the other is the requirement that
the Higgs potential does not lead to unrealistic minima.
We used the analytic expression for unrealistic minima obtained in Ref.~\cite{Kanehata:2011ei},
and checked whether the depth of the desired minimum with
$(\langle S\rangle,\langle H_u\rangle,\langle H_d\rangle)=(v_s/\sqrt{2},v_u/\sqrt{2},v_d/\sqrt{2})$
is deeper than that of possible unrealistic minima for given values of the parameters $(\lambda,\kappa)$.\footnote{An additional constraint might be imposed
in order to avoid charged and colored vacua~\cite{Ellwanger:1996gw}.
We do not put this constraint explicitly in Fig.~\ref{fig2}, since it also depends on other MSSM parameters such as soft masses for sleptons and squarks.}

%\if0

%\fi

The increase in the peak amplitude of gravitational waves for $\lambda\ll1$ is
straightforwardly understood from the relation in the decoupling limit [Eq.~\eqref{sigma_decoupling_limit}].
This relation implies that $\Omega_{\rm gw}h^2(t_0)_{\rm peak} \propto \kappa^2v_s^6t_{\rm dec}^2\propto\kappa^2\lambda^{-6}\mu^6t_{\rm dec}^2$
in the limit $\lambda\to 0$.\footnote{Note that the coupling $\lambda$ cannot be arbitrary small with keeping the relation in the decoupling limit [Eq.~\eqref{decoupling_limit}],
as was discussed at the end of Sec.~\ref{sec2}. This fact implies that we cannot extract arbitrarily large amounts of the energy of gravitational waves by simply keeping reducing the coupling $\lambda$.}
We see that the decoupling limit is relevant to the observations, since it generically predicts a large amplitude of gravitational waves.
Regarding this fact, in Table~\ref{tab1} we choose some bench mark points in the parameter space, whose predictions
can be tested in future experimental research.

\begin{table}[h]
\begin{center} %\scriptsize
\caption{Three bench mark points and estimated surface mass density of domain walls, peak amplitude of gravitational waves,
and its frequency. Here we used $g_*=10.75$ for $t_{\rm dec}=10^{-2}\mathrm{sec}$
and $g_*=68.8$ for $t_{\rm dec}=10^{-6}\mathrm{sec}$~\cite{Wantz:2009it} to estimate $\Omega_{\rm gw}h^2(t_0)_{\rm peak}$.}
\vspace{3mm}
\begin{tabular}{l l l l}
\hline\hline
$$ & I & II & III \\
\hline
$\lambda$ & $5\times 10^{-4}$ & $5\times 10^{-3}$ & $5\times 10^{-6}$ \\
$\kappa$ & $2\times10^{-4}$ & $2\times10^{-3}$ & $2\times10^{-6}$ \\
$A_{\lambda}$ & 150GeV & 150GeV & 150GeV \\
$A_{\kappa}$ & $-150$GeV & $-150$GeV & $-150$GeV \\
$\tan\beta$ & 5 & 5 & 5 \\
$\mu$ & 200GeV & 200GeV & 200GeV\\
$t_{\rm dec}$ & $10^{-2}\mathrm{sec}$ & $10^{-2}\mathrm{sec}$ & $10^{-6}\mathrm{sec}$ \\
\hline\hline
$\sigma_{\rm wall}$ & $1.96\times 10^4$ TeV$^3$ & $1.96\times 10^2$ TeV$^3$ & $1.96\times 10^8$ TeV$^3$ \\
$\Omega_{\rm gw}h^2(t_0)_{\rm peak}$ & $4.66\times 10^{-9}$ & $4.66\times 10^{-13}$ & $2.51\times 10^{-9}$ \\
$f(t_0)_{\rm peak}$  & $1.02\times 10^{-9}$Hz & $1.02\times 10^{-9}$Hz & $1.02\times 10^{-7}$Hz \\
\hline\hline
\label{tab1}
\end{tabular}
\end{center}
\end{table}

There are various ongoing and planned gravitational wave experiments, which aim to detect stochastic gravitational wave backgrounds as well as astrophysical sources.
Among them, ground-based interferometers such as (Advanced) LIGO~\cite{Abramovici:1992ah}, VIRGO~\cite{Acernese:2008zzf}, KAGRA~\cite{Somiya:2011np}, and ET~\cite{Sathyaprakash:2012jk}
will probe frequency range around $f\sim 100\mathrm{Hz}$.
Space-borne interferometers such as eLISA~\cite{AmaroSeoane:2012km} and DECIGO~\cite{Kawamura:2006up} are also planned to probe lower frequency ranges around 0.1mHz to 10Hz.
In addition to these experiments, gravitational wave backgrounds can be probed by using the Pulsar Timing Array (PTA)~\cite{Hobbs:2009yy}, which is sensitive to the frequency range
around $10^{-9}\textendash 10^{-8}\mathrm{Hz}$.
Recently, three projects including PPTA, EPTA, and NANOGrav reported upper limit on the stochastic gravitational
wave background of $\Omega_{\rm gw}h^2<\mathcal{O}(1)\times 10^{-8}$~\cite{Jenet:2006sv}.
As shown in Fig.~\ref{fig2}, this present upper limit already excludes some parameter region for $t_{\rm dec}=0.01$sec.
The future PTA observations such as FAST~\cite{Nan:2011um} and SKA~\cite{Cordes:2005gp} will further improve the sensitivities.

\begin{figure*}[htbp]
\begin{center}
\includegraphics[scale=1.0]{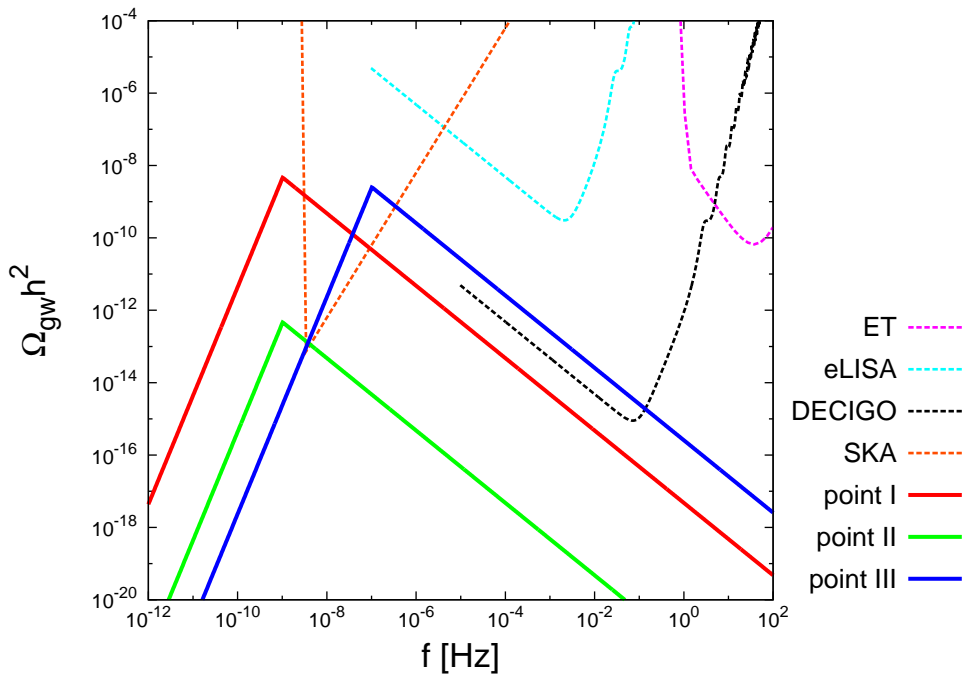}
\end{center}
\caption{Sensitivity curves of planned detectors (dotted lines) and the spectrum of the gravitational waves produced from domain walls in the NMSSM (solid lines)
for various choices of parameters (points I, II, and III) given in Tabe~\ref{tab1}.
The sensitivity curve for ET is based on~\cite{ET}, and those for eLISA and (Ultimate) DECIGO are based on~\cite{sens,Alabidi:2012ex} [see also Ref.~\cite{Hiramatsu:2013qaa}].
The sensitivity of SKA is taken from Ref.~\cite{Sesana:2010mx}.
}
\label{fig3}
\end{figure*}

Figure~\ref{fig3} shows the schematics of the spectrum of gravitational waves predicted from three points enumerated in Table~\ref{tab1} and sensitivities of planned experiments.
The recent result of numerical simulations~\cite{Hiramatsu:2013qaa} indicates that the spectrum of gravitational waves from domain walls scales like $\Omega_{\rm gw}\propto f^3$
for $f<f_{\rm peak}$ and $\Omega_{\rm gw}\propto f^{-1}$ for $f>f_{\rm peak}$, and we also plot this expectation in the figure.
We see that the point I leads to the peak at the frequency $f \sim 10^{-9}\mathrm{Hz}$, which can be probed in PTA observations such as SKA.
On the other hand, the prediction of the point II indicates that the amplitude gets smaller as the values of $\lambda$ and $\kappa$ increase.
Interestingly, the spectrum of gravitational waves predicted from the point III might be probed by DECIGO as well as SKA.
From these illustrations, we expect that
the future PTA observations and space-borne interferometers can probe the signature of gravitational waves from NMSSM domain walls.

%%%%%%%%%%%%%%%%%%%%%%%%%%%%%%%%%%%%%%%%%%%%%%%%%%%%%%%%%%%%%%%%
\section{\label{sec5} Conclusions}
%%%%%%%%%%%%%%%%%%%%%%%%%%%%%%%%%%%%%%%%%%%%%%%%%%%%%%%%%%%%%%%%
In this paper, we estimated the gravitational wave signatures from domain walls created at the electroweak phase transition in the NMSSM.
The surface mass density of domain walls $\sigma_{\rm wall}$ is estimated by solving nonlinear equations for Higgs fields numerically.
By using the estimated value of $\sigma_{\rm wall}$, the amplitude of relic gravitational waves is calculated via Eq.~\eqref{Omega_gw_t0_peak_2}
and its parameter dependence is explored. We find that a large amplitude of gravitational waves, which scales as $\Omega_{\rm gw}h^2_{\rm peak}\propto \kappa^2v_s^6 t_{\rm dec}^2$,
is predicted in the decoupling limit. Furthermore, the peak frequency is determined by the decay time of domain walls [Eq.~\eqref{f_t0_peak_2}],
which is related to the magnitude of the bias term $\Delta V$ via Eq.~\eqref{t_dec}.
Such gravitational wave signatures can be probed in the future experiments with improved sensitivities.

We note that the decay time of domain walls $t_{\rm dec}$ is restricted to a certain interval, from the epoch of the electroweak phase transition to that of BBN.
The typical frequency of gravitational waves corresponding to such time scales is very low, which is suitable for PTA observations.
Since the amplitude of gravitational waves becomes large in the decoupling limit, we can use the results of PTA observations
as a lower limit on the couplings ($\lambda$ and $\kappa$) appearing in the Higgs sector of the NMSSM.
For instance, the present PTA bound $\Omega_{\rm gw}h^2\lesssim \mathcal{O}(10^{-8})$
already puts a lower limit $\lambda>10^{-4}\textendash 10^{-3}$ for $t_{\rm dec}=0.01\mathrm{sec}$ (see Fig.~\ref{fig2}),
while this constraint becomes irrelevant if the decay time of domain walls $t_{\rm dec}$ is earlier than $\mathcal{O}(0.01)\mathrm{sec}$.
Future observations will probe a wider range of the parameters in the weak coupling regime,
which can be used to complement the results of other experimental studies of the NMSSM.
%%%%%%%%%%%%%%%%%%%%%%%%%%%%%%%%%%%%%%%%%%%%%%%%%%%%%%%%%%%%%%%%
\begin{acknowledgments}
This work was supported by Grant-in-Aid for Scientific research from the Ministry of Education,
Science, Sports, and Culture (MEXT), Japan, No. 25400248 (M.~K.), World Premier International Research Center Initiative (WPI Initiative), MEXT, Japan.
Numerical computation in this work was carried out at the Yukawa Institute Computer Facility.
K.~K.~is supported by Institute for Basic Science (IBS-R018-D1).
K.~S.~is supported by the Japan Society for the Promotion of Science through research fellowships.
\end{acknowledgments}
%%%%%%%%%%%%%%%%%%%%%%%%%%%%%%%%%%%%%%%%%%%%%%%%%%%%%%%%%%%%%%%%

%%%%%%%%%%%%%%%%%%%%%%%%%%%%%%%%%%%%%%%%%%%%%%%%%%%%%%%%%%%%%%%%

\end{document}